# An Empirical Investigation on Important Subgraphs in Cooperation-Competition networks

A.-X. Feng, C.-H. Fu, X.-L. Xu, Ai-Fen Liu, H. Chang, D.-R. He and G.-L. Feng

College of Physics Science and Technology, Yangzhou University, Yangzhou, 225002, China

fax20032008@gmail.com

**Abstract**

Subgraphs are very important for understanding structure and function of complex networks. Dyad and triad are the elementary subgraphs. We focus on the distribution of their act degree defined as the number of activities, events or organizations they join, which indicates the importance of the subgraphs. The empirical studies show that, in a lot of real world systems, the dyad or triad act degree distributions follow "shifted power law" (SPL), $P(q) \propto (q+a)^{-g}$, where $\alpha$ and $\gamma$ are constants. We defined a "heterogeneity index", $H$, to describe how it is uneven and analytically deduced the correlation between $H$ and $\alpha$ and $\gamma$. This manuscript, which shows the details of the empirical studies, serves as an online supplement of a paper submitted to a journal.

**1. Introduction**

Complex networks attract research interests recently [1, 2, 3]. Among the studies, social collaboration networks [4-7] and cooperation-competition networks [8-17] become important topics. An important type of the networks is described by bipartite graphs, in which vertices can be divided into two sets [18]. One type of the vertices is "actors" taking part in some activities, organizations or events, which can be viewed as another type of vertices called "acts". The links only exist between different set of vertices.

In network, community is an important quantity [18]. It can be divided into some subgraphs. The subgraphs contain several actors and the connections among them. It is of importance to know subgraphs. As an example, in biological networks, subgraphs often relate to functional motifs [19-21]. In social networks, subgraphs often are composed of basic social units, such as husband and wife, husband wife and son, best friends and so on. The smallest subgraph is composed of a pair of actors and the edge between them, called "dyad" in sociology [18]. Similarly, "triad" means a triple of actors and the edges among them. Here we only consider mutual dyads and triads where the edge connecting them is un-directed. In Ref. [8] Zhang et al. addressed "act degree" as the number of acts an actor takes part in. Here we extend it to the number of acts a dyad or triad joins. The act degree distribution of dyad (triad) describes the dyads' (triads') overall information and becomes the most important property in our current study. The number of acts, which a dyad (triad) takes part in, is addressed as "dyad act degree" (denoted by $D$) (triad act degree denoted by $T$). The distribution $P(D)$ or $P(T)$ is defined as the probability with which a dyad or triad takes part in a certain number of acts, $D$ or $T$. The dyad (triad) act degree definition can be more accurately expressed as $D_{i,j} = \sum_m a_{i,m} a_{j,m}$ and $T_{i,j,k} = \sum_m a_{i,m} a_{j,m} a_{k,m}$ respectively where $i, j, k$ denote different actors and $m$ means an act. $a_{i,m}$ denotes the element of adjacency matrix. If actor $i$ takes part in act $m$, $a_{i,m} = 1$, otherwise $a_{i,m} = 0$. We use $d$ and $t$ to denote the normalized $D$ and $T$ respectively so that $P(d)$ and $P(t)$ express the normalized distributions.

From empirical investigations (as presented in the following sections) we know that the dyad and triad act degree distributions follow the so-called "shifted power law" (SPL), $P(q) \propto (q+a)^{-g}$, where $\alpha$ and $\gamma$ are constants [8,9]. The "SPL" function interpolates between a power law and an exponential function, when α varies from 0 to 1. Another parameter, heterogeneity index, can be used to character the uniformity or inequality of act degree distributions. Recently Hu and Wang have applied it to characterize the heterogeneity of network degree distribution [22, 23], and Xu et al employed it to describe heterogeneity of network weight distribution [15,17]. To define the heterogeneity index, the normalized act degree of each dyad (triad) is sorted by increasing order, i.e. $d_1 \leq d_2 \leq \cdots \leq d_N$ or $t_1 \leq t_2 \leq \cdots \leq t_N$ ($N$ is the total number of dyads or triads). An *x-y* plane is defined where $x_i = i/N$ and $y_i = \sum_j^i d_j$ (or $y_i = \sum_j^i t_j$ ). The disparity of $d$ (or $t$) value distribution can be shown by the $y(x)$ line on the $x - y$ plane. As shown in Fig. 1-1, the heterogeneity index $H$ can be dened as $H = S_A/(S_A + S_B) = 1 - 2 S_B$ [23] where $S_A$ denotes the area between the diagonal line, $x = y$ (which shows that the distribution is absolutely homogeneous), and the $y(x)$ line; and $S_B$ denotes the area beneath the $y(x)$ line, namely the area between the $y(x)$ line, $y = 0$ axis, and $x = 1$ axis (the two axes indicate that the distribution is completely heterogeneous since $d$ (or $t$) is concentrated on only one value).

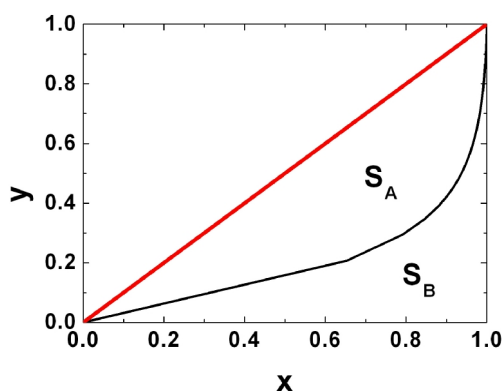

Fig. 1-1   A schematic showing the definition of $H$.

We empirically investigated the dyad (triad) act degree distributions and their heterogeneity indexes in 13 real world networks. In this online supplement we present the empirical investigation details. In each of the following sections we present the empirical investigation results obtained in one real-world system.

## 2. Chinese training organization

In order to enter a better high school, university or find a better job, a lot of students in China choose a training organization for the competition. The actors are defined as the training organizations and the acts are defined as the training courses the organizations offer. If two training organizations offer at least one course, they are connected by an edge and form a dyad denoting their competition in the training market. The act degree of the dyad is defined as the number of the courses, which are offered by the two organizations. It signifies how severe the competition is. The definition is similar for a triad (and the triad act degree) but it contains three organizations. The data of the Chinese training organizations network are downloaded from http://www.ot51.com, http://www.00100.cc and http://www.peopele.com.cn. 2674 training courses and 398 training organizations are included in the data. The empirical investigations on the dyad (triad) act degree distributions and their heterogeneity indexes are shown by the following four figures. The distributions can be well fitted by power laws, which are special forms of SPL. The parameter values, $H$ and $α$ and $γ$, will be presented by a table at the end of the manuscript.

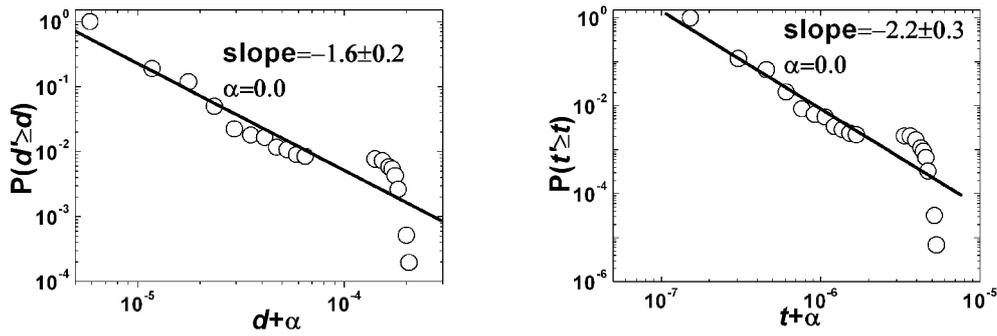

Fig.2-1 (left). Cumulative dyad act degree distribution of training organization; and Fig.2-2 (right). Triad act degree distribution of training organization

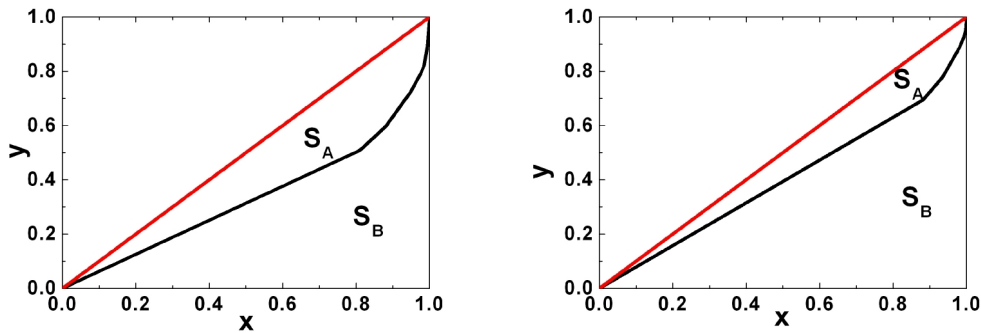

Fig.2-3 (left). Empirical investigation on dyad act degree distribution heterogeneity index; and Fig.2-4 (right). On triad act degree distribution heterogeneity index of training organization

### 3. Chinese university recruit

In China, all the universities can recruit new students from middle schools only via a nationwide uniform activity. All the middle students have to pass the national matriculation. The delegacies of different academic level universities then get together to select middle school students according to the matriculation marks of the students and the university batch rights. However, in recent years, some best universities are awarded the right to recruit middle school students before the national matriculation. They can perform their own examinations or interviews and then recruit new students by their own decision. We call this as "independent recruitment". The middle schools can achieve better reputations if more of their students have been recruited by these top level universities. Therefore, the middle schools collaborate to accomplish the independent recruitment activity and simultaneously compete for sending more students to the universities. We define the middle schools as the actors, the independent recruitment universities as the acts. In 2006, there were 52 universities, which had the independent recruitment right. 1546 middle schools were included in the data (from www.chsi.com.cn). The empirical investigations on the dyad (triad) act degree distributions and their heterogeneity indexes are shown by the following four figures. The distributions can be well fitted by SPL functions.

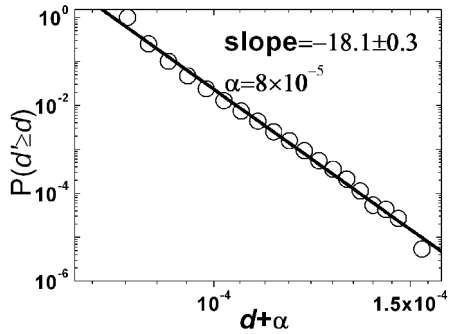 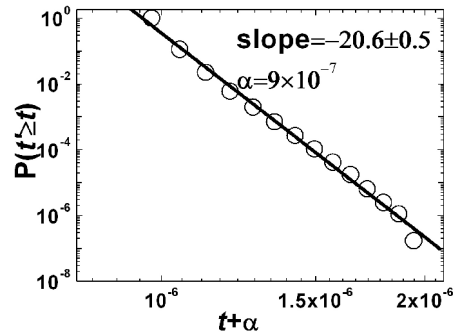

Fig.3-1 (left). Cumulative dyad act degree distribution of Chinese university recruit; and Fig.3-2 (right). Triad act degree distribution of Chinese university recruit

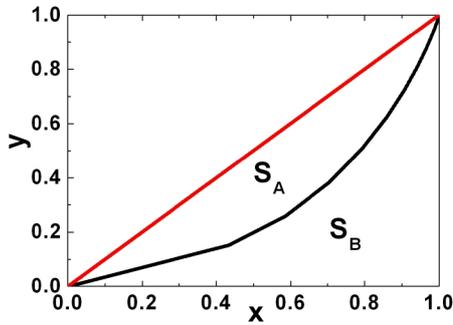 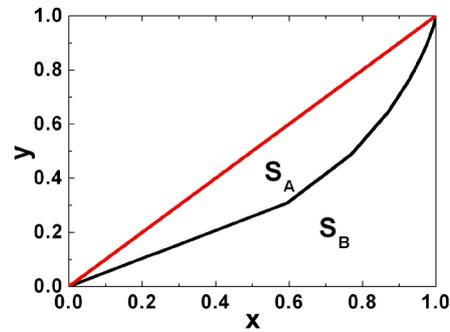

Fig.3-3 (left). Empirical investigation on dyad act degree distribution heterogeneity index; and Fig.3-4 (right). On triad act degree distribution heterogeneity index of Chinese university recruit

## 4. World language distribution

We propose a network description on world language distribution. The actors are defined as languages. The acts are defined as countries or regions where the languages are spoken. Two actors are connected by an edge if they are coexisting in at least one common region. In a very long time consideration, the languages can be considered as being competing to be used by more people. As the result, some languages have died out, but some other languages have been spoken by more and more people and spread to more and more geographical regions. The languages also collaborate in the common regions to accomplish the communications between the people. The data were downloaded from the 15th edition of Ethnologue (http://www.ethnologue.com), published in 2005, which lists 6142 languages and 228 countries plus 8 regions. The empirical investigations on the dyad (triad) act degree distributions and their heterogeneity indexes are shown by the following four figures. Note that this system shows a very small $S_A$ that induces an almost zero heterogeneity index.

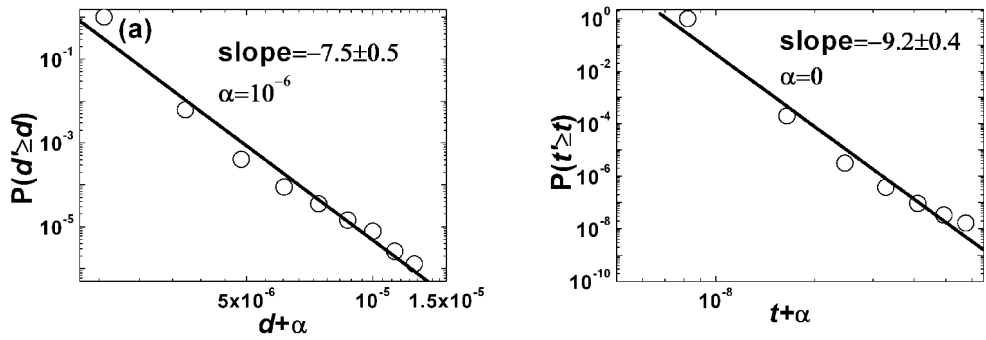

Fig.4-1 (left). Cumulative dyad act degree distribution of world language distribution; and Fig.2-2 (right). Triad act degree distribution of world language distribution

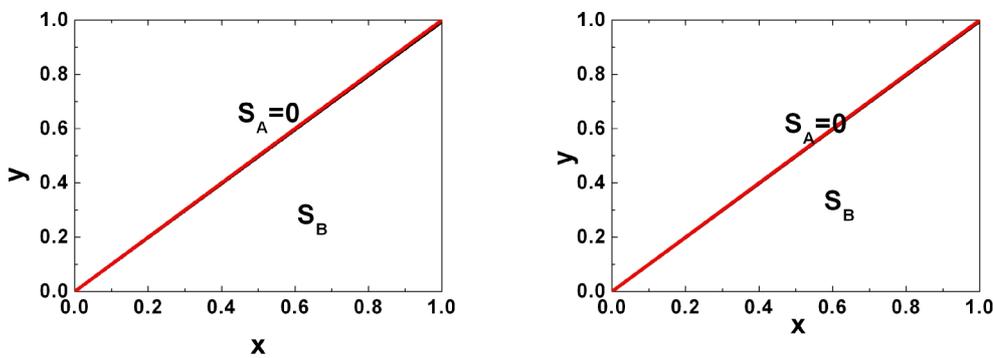

Fig.4-3 (left). Empirical investigation on dyad act degree distribution heterogeneity index; and Fig.4-4 (right). On triad act degree distribution heterogeneity index of world language distribution

## 5. Human acupuncture points

Acupuncture is one of the most important traditional Chinese therapy means. Because of its practical efficacy in curing some diseases, acupuncture becomes more and more popular over the world. There are 187 key acupuncture points in human body. To treat a certain disease, several specific acupuncture points need to be acupunctured simultaneously. In this sense, the acupuncture points collaborate in curing a certain disease. Therefore, we can define the diseases treated by acupuncture as acts, and the acupuncture points as actors. Several actors are linked if they are used in curing at least one disease. Totally 108 different kinds of diseases and 187 acupuncture points are involved in our data (from www.acutimes.com ). The empirical investigations on the dyad (triad) act degree distributions and their heterogeneity indexes are shown by the following four figures.

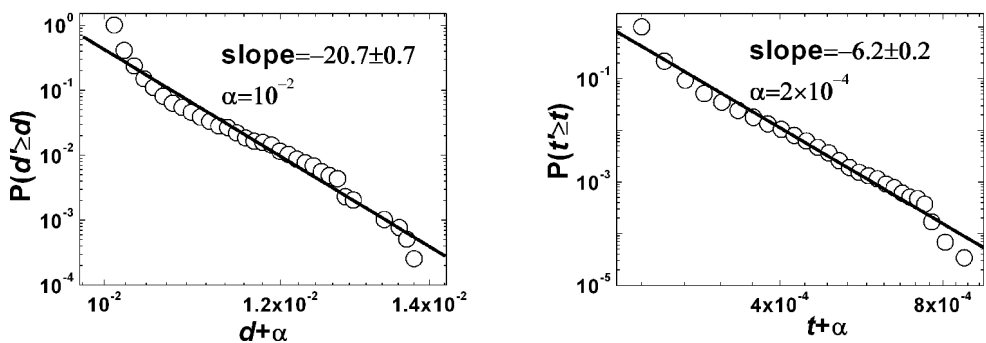

Fig.5-1 (left). Cumulative dyad act degree distribution of acupuncture points; and Fig.5-2 (right). Triad act degree distribution of acupuncture points

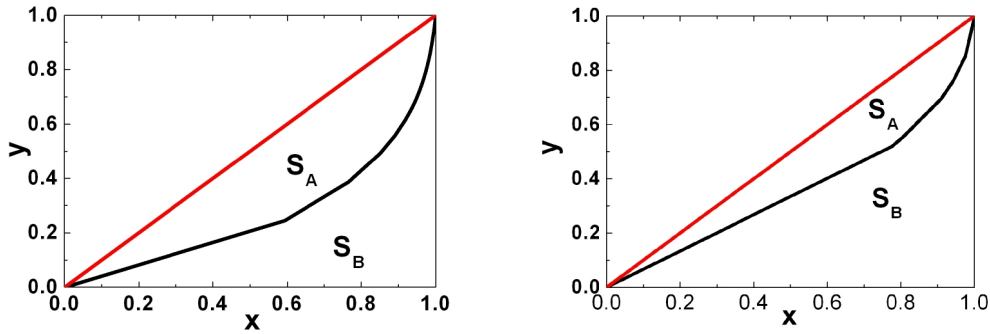

Fig 5-3 (left). Empirical investigation on dyad act degree distribution heterogeneity index; and Fig.5-4 (right). On triad act degree distribution heterogeneity index of acupuncture points

## 6. Herb prescription formulation

Chinese herbology is a result from the simple dialectical materialism philosophy of the traditional Chinese. It emphasizes the inner cause of any illness and states that a healthy body should be able to maintain a dynamic balance against the world outside of the body through self-adjustment. Therefore, any illness is due to unbalance between inner self and outside world. However, the balance is affected by many different and complicated changing factors, and unbalance cannot always be treated by a single herb, which is able to attack just a few problems. In addition, any herb has side effects. It may be effective in dealing with one factor but it would cause other unbalance. So we need to combine appropriate herbs to make an effective prescription where different herbs work together in a complementary manner in order to cure an illness and minimize the side effects. We define every herb as an actor, and draw a link between herbs included in a prescription (defined as acts) to represent their interactions, i.e., their collaboration relationship at the process of curing an illness. We have included 1536 prescriptions and 681 herbs (actors) from Refs. [24,25] for our study. These herbal prescriptions are the results of long-term experiments conducted by the Chinese people and serve as the representative samples of the prescription population. The empirical investigations on the dyad (triad) act degree distributions and their heterogeneity indexes are shown by the following four figures.

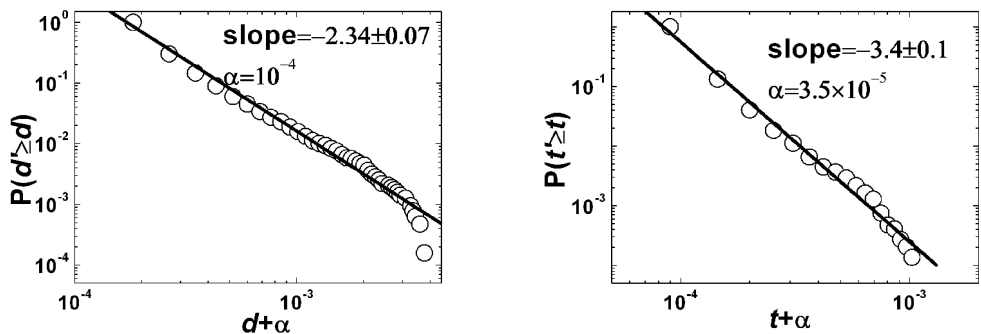

Fig.6-1 (left). Cumulative dyad act degree distribution of Chinese herbology; and Fig.6-2 (right). Triad act degree distribution of Chinese herbology

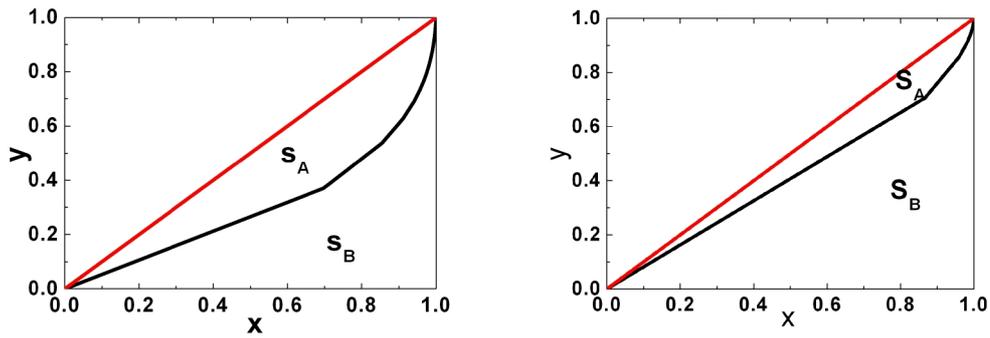

Fig 6-3 (left). Empirical investigation on dyad act degree distribution heterogeneity index; and Fig.6-4 (right). On triad act degree distribution heterogeneity index of Chinese herbology

## 7. Software downloading

In the software downloading network an actor is defined as software and the acts are defined as the web sites where the software is downloaded. If two actors are downloaded from at least one common web site, they are connected by an edge and form a dyad. The act degree of the dyad is defined as the number of the web sites, which offer the software. The definition is similar for a triad (and the triad act degree) but it contains three actors. The data are downloaded from http://www.baidu.com/. 3844 software and 128 web sites are included in the data. The empirical investigations on the dyad (triad) act degree distributions and their heterogeneity indexes are shown by the following four figures. The distributions can be well fitted by power laws, which are special forms of SPL.

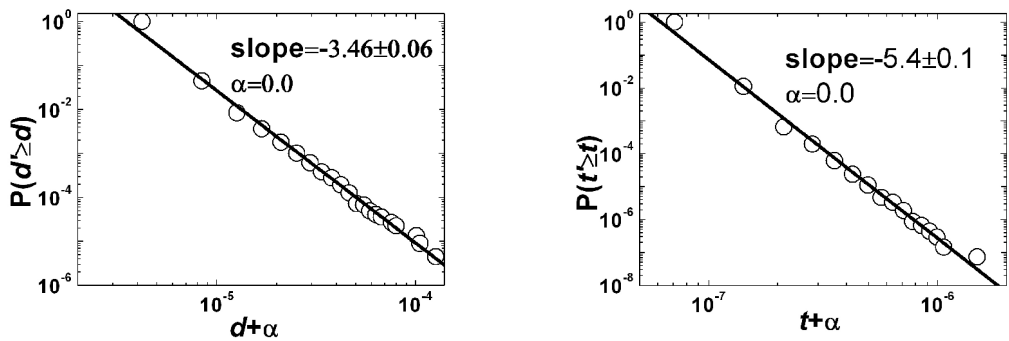

Fig.7-1 (left). Cumulative dyad act degree distribution of software downloading; and Fig.7-2 (right). Triad act degree distribution of software downloading

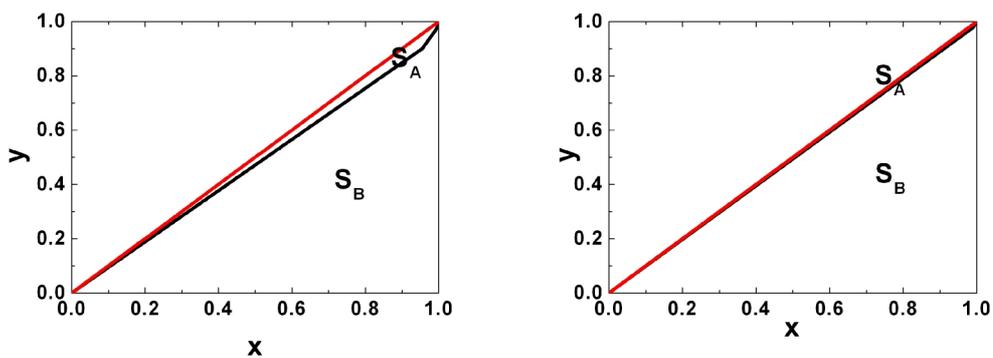

Fig 7-3 (left). Empirical investigation on dyad act degree distribution heterogeneity index; and Fig.7-4 (right). On triad act degree distribution heterogeneity index of software downloading

## 8. IT product market

The market of information technique (IT) products, including mobile phone, computer, digital camera, and so on, is another example system. If different manufacturers produce the same type of IT products, they compete in the selling market. Of course, these manufacturers also collaborate to supply enough IT products, and to maintain the market order. We construct a collaboration-competition bipartite network in which the manufactures are defined as actors and the IT products are defined as acts. On the website www.pcpop.com, there are detailed introductions to each IT product produced by a specific manufacturer. We collected 265 manufacturers and 2121 IT products from the website. The empirical investigations on the dyad (triad) act degree distributions and their heterogeneity indexes are shown by the following four figures.

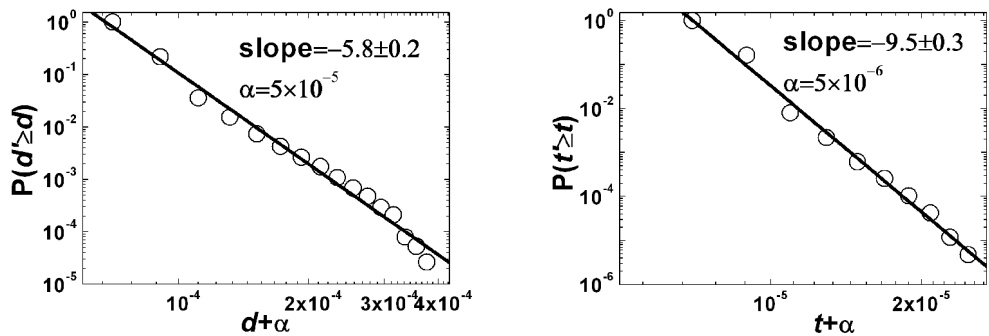

Fig.8-1 (left). Cumulative dyad act degree distribution of IT product market; and Fig.8-2 (right). Triad act degree distribution of IT product market

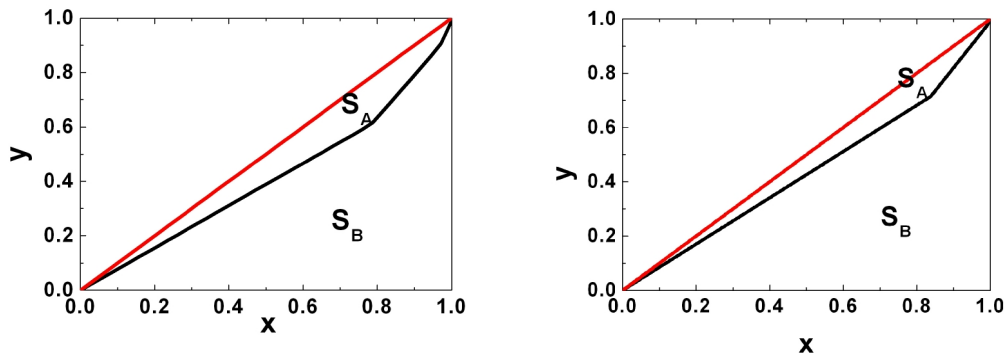

Fig 8-3 (left). Empirical investigation on dyad act degree distribution heterogeneity index; and Fig.8-4 (right). On triad act degree distribution heterogeneity index of IT product market

## 9. Olympic game (2004)

In an Olympic game some athletes join a sport event to successfully conduct the pageant and also to obtain more sport scores. We construct the network of the 2004 Athens Olympic Game by defining the athletes as the actor nodes and the sport events (only the individual sport events, e.g., high jump, weight lifting, are considered) as acts. The data were downloaded from www.sina.com.cn (2004), which includes 133 individual sport events and 4500 athletes, as well as their sport scores in each sport event. An edge in the bipartite graph represents that an athlete takes part in an event. In the projected unipartite graph two actors are connected if they join in at least one common act. The empirical investigations on the dyad (triad) act degree distributions and their heterogeneity indexes are shown by the following four figures.

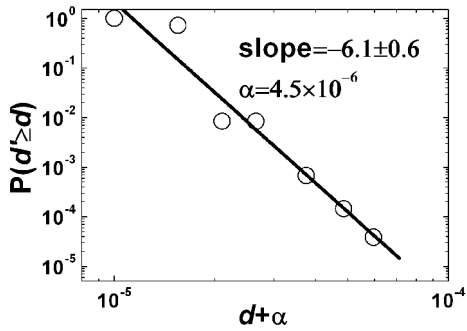 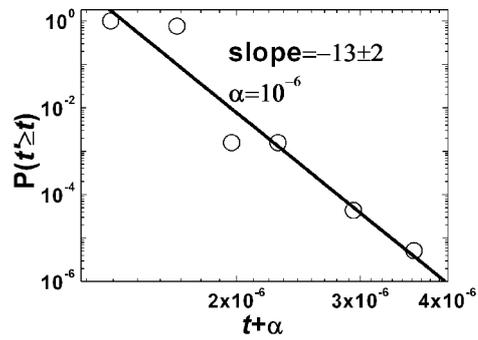

Fig.9-1 (left). Cumulative dyad act degree distribution of 2004 Athens Olympic Game; and Fig.9-2 (right). Triad act degree distribution of 2004 Athens Olympic Game

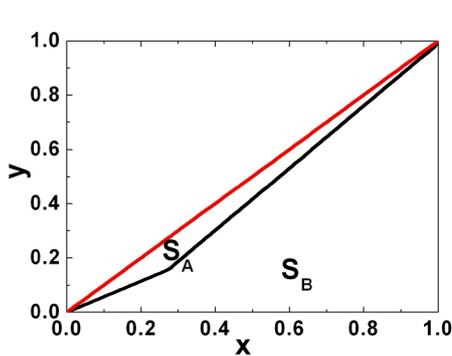 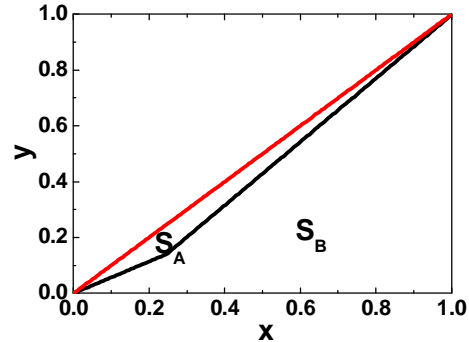

Fig 9-3 (left). Empirical investigation on dyad act degree distribution heterogeneity index; and Fig.9-4 (right). On triad act degree distribution heterogeneity index of 2004 Athens Olympic Game

## 10. Huai-Yang recipes

We choose 329 recipes of the Huai-Yang system (Huai-Yang denotes two geographical regions located in the middle-eastern part of China) of Chinese cooked food, which are defined as acts; there are a total of 242 foods defined as actors. Two actors are connected by an edge if they form at least one recipe. The data are collected from Ref. [26]. The empirical investigations on the dyad (triad) act degree distributions and their heterogeneity indexes are shown by the following four figures.

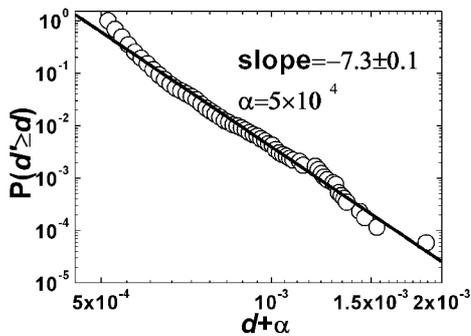 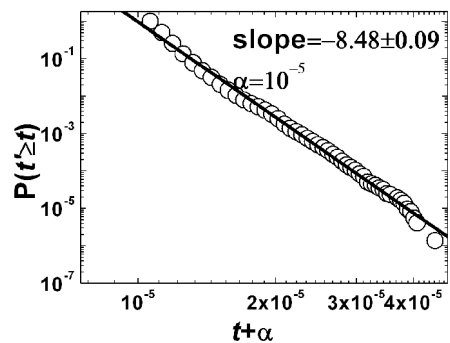

Fig.10-1 (left). Cumulative dyad act degree distribution of Huai-Yang recipes; and Fig.10-2 (right). Triad act degree

distribution of Huai-Yang recipes

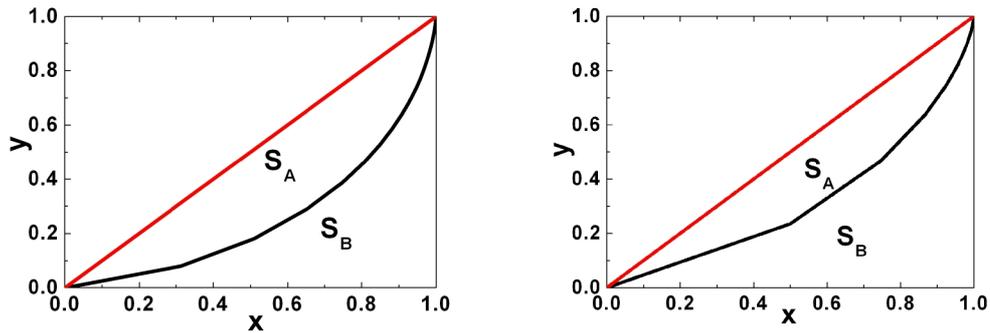

Fig 10-3 (left). Empirical investigation on dyad act degree distribution heterogeneity index; and Fig.10-4 (right). On triad act degree distribution heterogeneity index of Huai-Yang recipes

## 11. Mixed drinks

The mixed drinks (such as cocktails) usually contain a large number of ingredients according to the consumer's taste, and many mixed drinks may share the same ingredients. We construct the mixed drink network by defining the component ingredients as actors and the mixed drinks as acts. The ingredients collaborate to form mixed drinks with different tastes. Simultaneously, the ingredients contained in a common mixed drink can be regarded as being competing since the ingredients have different relative importance. Until 2006, we collected 7804 mixed drinks and 1501 ingredients from www.drinknation.com. The empirical investigations on the dyad (triad) act degree distributions and their heterogeneity indexes are shown by the following four figures.

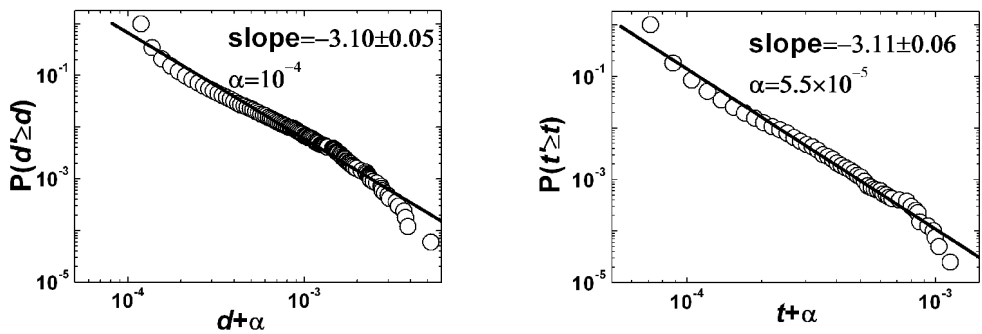

Fig.11-1 (left). Cumulative dyad act degree distribution of mixed drinks; and Fig.11-2 (right). Triad act degree distribution of mixed drinks

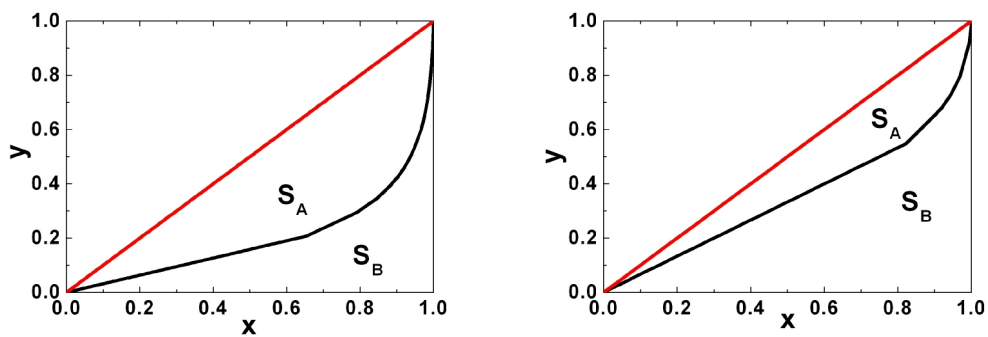

Fig 11-3 (left). Empirical investigation on dyad act degree distribution heterogeneity index; and Fig.11-4 (right). On triad act degree distribution heterogeneity index of mixed drinks

## 12. Beijing bus route

The study of traffic networks is always of great interest. We define the bus route in Beijing city as an act, and a station as an actor. The edge between two actors expresses the collaboration in a common route of two actors. 572 bus routes and 4199 bus stops (in 2006) were collected from http://www.bjbus.com/. The empirical investigations on the dyad (triad) act degree distributions and their heterogeneity indexes are shown by the following four figures.

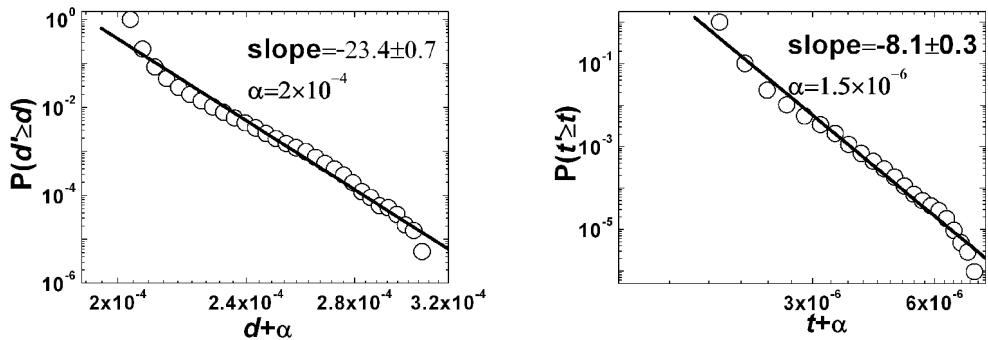

Fig.12-1 (left). Cumulative dyad act degree distribution of Beijing bus route; and Fig.12-2 (right). Triad act degree distribution of Beijing bus route

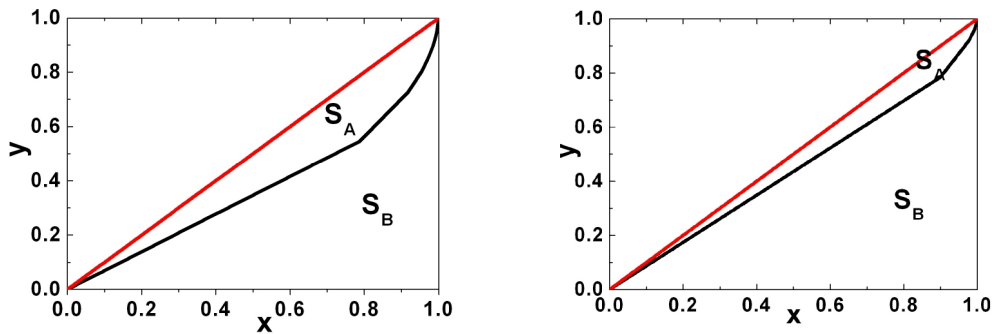

Fig 12-3 (left). Empirical investigation on dyad act degree distribution heterogeneity index; and Fig.12-4 (right). On triad act degree distribution heterogeneity index of Beijing bus route

## 13. PC Sales on Taobao

In recent years, on-line shopping by internet becomes more and more popular. Taobao (www.taobao.com) is one of the most famous on-line shopping mall in China. Many shops sell many kinds of commodities through the Taobao website. Thousands of shops sell notebook PCs. The shops collaborate to provide proper notebook PC selling service, and simultaneously compete for more profit. In this bipartite network, the shops are defined as actors, and the selling markets of the notebook PC types are defined as the acts. Totally 53 notebook PC types and 4711 notebook PC shops in the Taobao on-line shopping mall were collected. The empirical investigations on the dyad (triad) act degree distributions and their heterogeneity indexes are shown by the following four figures.

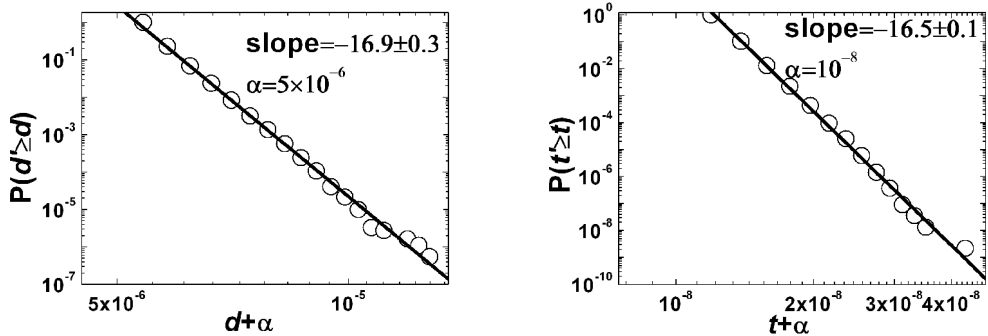

Fig.13-1 (left). Cumulative dyad act degree distribution of PC Sales on Taobao; and Fig.13-2 (right). Triad act degree distribution of PC Sales on Taobao

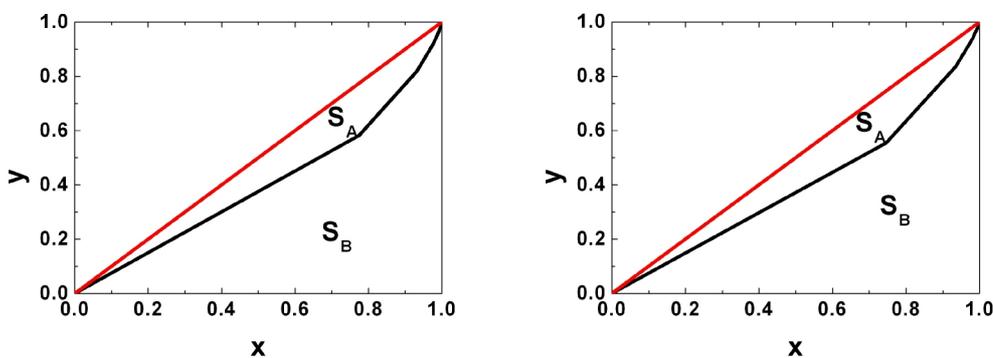

Fig 13-3 (left). Empirical investigation on dyad act degree distribution heterogeneity index; and Fig.13-4 (right). On triad act degree distribution heterogeneity index of PC Sales on Taobao

## 14. The travel route of China

Chinese travel route network is set up by defining the travel routes as acts and the scenic spots as the actors. Two scenic spots are linked if at least one travel route contain both of them. The system consists of 219 scenic spots and 190 travel routes (http://www.cnta.com/8-ssls/lyqd.asp/ ). The empirical investigations on the dyad (triad) act degree distributions and their heterogeneity indexes are shown by the following four figures.

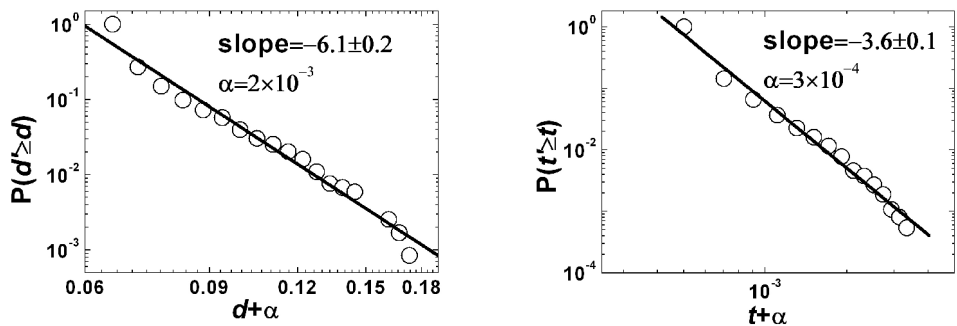

Fig.14-1 (left). Cumulative dyad act degree distribution of travel route of China; and Fig.14-2 (right). Triad act degree distribution of travel route of China

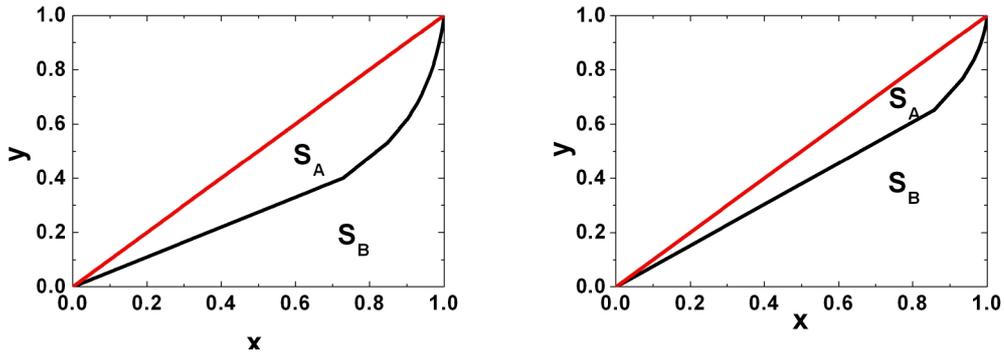

Fig 14-3 (left). Empirical investigation on dyad act degree distribution heterogeneity index; and Fig.14-4 (right). On triad act degree distribution heterogeneity index of travel route of China

As a summary, Table 1 lists all the parameter values of these 13 real world systems.

Table 1: $N_1$ or $N_2$ means the total number of dyads (2-clique) or triads(3-clique), respectively; $a_1$ or $a_2$ (and $g_1$ or $g_2$) denote the two parameters of SPL fitting functions for dyad or triad act degree distributions, respectively (since the figures in the online supplement and this section show cumulative distributions, the values of $g$ listed here are equal to the slopes (shown in the figures) plus one); $H_1$ or $H_2$ is the heterogeneity index of dyad or triad act degree, respectively.

| Real world networks | $N_1$ | $N_2$ | $a_1$ | $a_2$ | $g_1$ | $g_2$ | $H_1$ | $H_2$ |
|---|---|---|---|---|---|---|---|---|
| Chinese training organization | 106595 | 5163310 | 0.0 | 0.0 | 2.6 | 3.2 | 0.34 | 0.20 |
| Software downloading | 223354 | 13919001 | 0.0 | 0.0 | 4.5 | 6.4 | 0.05 | 0.01 |
| Olympic game (2004) | 103974 | 1764550 | $4.5\times 10^{-6}$ | $1.0\times 10^{-6}$ | 7.1 | 14.0 | 0.12 | 0.11 |
| Herb prescription formulation | 6362 | 14794 | $1.0\times 10^{-4}$ | $3.5\times 10^{-5}$ | 3.3 | 4.4 | 0.40 | 0.17 |
| World living language | 771555 | 121661394 | $1.0\times 10^{-6}$ | 0.0 | 8.5 | 10.2 | 0.00 | 0.00 |
| Mixed drinks | 16881 | 40222 | $1.0\times 10^{-4}$ | $5.5\times 10^{-5}$ | 4.1 | 4.1 | 0.60 | 0.30 |
| IT product | 38083 | 429640 | $5.0\times 10^{-5}$ | $5.0\times 10^{-6}$ | 6.8 | 10.5 | 0.18 | 0.13 |
| The travel route of China | 1188 | 3754 | $2.0\times 10^{-3}$ | $3.0\times 10^{-4}$ | 7.1 | 4.6 | 0.39 | 0.22 |
| Huai-Yang recipes | 17278 | 733956 | $5.0\times 10^{-4}$ | $1.0\times 10^{-5}$ | 8.3 | 9.5 | 0.49 | 0.37 |
| PC Sales on Taobao | 1839208 | 461272179 | $5.0\times 10^{-6}$ | $1.0\times 10^{-8}$ | 17.9 | 17.5 | 0.21 | 0.21 |
| Chinese university recruit | 187352 | 11746963 | $8.0\times 10^{-5}$ | $9.0\times 10^{-7}$ | 19.1 | 21.6 | 0.26 | 0.11 |
| Beijing bus route | 192009 | 3186314 | $2.0\times 10^{-4}$ | $1.5\times 10^{-6}$ | 24.4 | 9.1 | 0.27 | 0.12 |
| Human acupuncture points | 3924 | 29706 | $1.0\times 10^{-2}$ | $2.0\times 10^{-4}$ | 21.7 | 7.2 | 0.47 | 0.29 |

## 15. Relationship between heterogeneity index and $a$, $g$

As mentioned, this manuscript serves as an online supplement of a journal paper, which reports an analytical discussion on the relationship between heterogeneity index and $a$, $g$. The conclusion is that the relationship between the heterogeneity index of dyad act degree and the key parameters $a$ and $g$ of dyad act degree distribution function (when $g$ is greater than 2 and the dyad number $N$ is close to infinity) is:

$$h_\infty = \frac{1+Na}{2g-3} \quad (g > 2)$$

It is the same for the triad. An agreement between the empirical results and this analytic conclusion was observed. The details will be published in the journal paper.

**Acknowledgment**

This work is supported by the State Natural Science Foundation of China (Grant NO. 40930952, 40875040, 40905034 and 10635040).

**References**

[1] R. Albert, A-L. Barabasi, Rev. Mod.Phys. 74 (2002)47.

[2] S.N. Dorogovtsev and J.F.F. Mendes, Adv. Phys. 51,1079(2002).

[3] M.E.J Newman, SIAM Review 45(2003)167.

[4] D.J. Watts, S.H. Strogatz, Nature 393(1998)440.

[5] A-L. Barabasi, R. Albert, Science 286(1999)509.

[6] M.E.J. Newman, Phys. Rev. E 64, 016131 (2001); M.E.J. Newman, Phys.Rev. E 64, 016132 (2001).

[7] J.J. Ramasco, S.N. Dorogovtsev, R. Pastot-Satorras, Phys. Rev. E 70, 036106 (2004).

[8] P.P. Zhang, K.Chen, Y. He, et al. Physica A 360(2006)599.

[9] H. Chang, B.-B. Su, Y.-P. Zhou, D.-R. He, Physica A 383 (2007) 687.

[10] B.B. Su, H. Chang, Y.-Z. Chen, D. R. He, Physica A 379 (2007) 291.

[11] C.-H. Fu, Z.-P. Zhang, H. Chang, et al., Physica A 387 (2008) 1411.

[12] C.-H. Fu, X.-L. Xu, A.-F. Liu, Y.-P. Wu, D. Shen, S.-J. Liu, X. Qian, Y.-C. Feng, C.-L. Wei, D.-R. He, Chin. Phys. Lett. 25, 4181 (2008).

[13] J.-J. Shi, Y.-L. Wang, C-H. Fu et al., Chin. Phys. Lett. 26, 078902 (2009).

[14] X.-L. Xu, C.-H. Fu, C-P. Liu, D-R. He, Chin. Phys. B. 19, 060501R (2010).

[15] X..-L. Xu, C.-H. Fu, A.-F. Liu, D.-R. He, Chin. Phys. Lett. 27, 048901 (2010).

[16] X..-L. Xu, C.-H. Fu, H. Chang, D.-R. He, An Evolution Model of Complex Systems with Simultaneous Cooperation and Competition, arxiv-1101.0406 (2010).

[17] X.-L. Xu,S.-R. Zou, C.-H. Fu, H. Chang, and D.-R. He, Cooperation sharing distributions in some cooperation-competition systems, arxiv:1010.0616 (2010)

[18] S. Wasserman, K. Faust, Social Network Analysis: Methods and Applications, Cambridge Univ. Press, Cambridge, 1994.

[19] R. Milo et al., Science 298 (2002)824.

[20] A.-L. Barabasi et al., Nat. Rev. Genet 5 (2004)101.

[21] S. Itzkovitz, R. Milo, N. Kashtan, G. Ziv, and U. Alon, Phys.Rev. E 68, 026127(2003).

[22] H.-B. Hu and L. Wang, Advances in complex systems, 8 (1) (2005) 159.

[23] H.-B. Hu and X. F. Wang , Physica A 387 (2008) 3769.

[24] Y. -X. Zhu, Guide to Chinese Medicine Prescription, Jindun: Beijing, 1996.

[25] D. -L. Liu et al, Prescription medicine commonly used manual, People's Medical: Beijing, 1996.

[26] The group of Minzu Hotel in Beijing, Huai-Yang Menu, China Tourism: Beijing, 1993.